\documentclass[a4paper,11pt]{article}
\usepackage{pos}
\usepackage{hyperref}
\usepackage{cleveref}
\usepackage{amsmath}
\usepackage{dsfont}
\usepackage{graphicx}

\title{Benchmark Continuum Limit Results for Spectroscopy with Stabilized Wilson Fermions}
\ShortTitle{Benchmark Continuum Limit Results for Spectroscopy with SWF}

\author[a]{F. Cuteri}
\author[b]{A. Francis}
\author[c]{P. Fritzsch}
\author*[d]{G. Pederiva}
\author[e]{A. Rago}
\author[f]{A.~Shindler}
\author[g]{A.~Walker-Loud}
\author[h]{S. Zafeiropoulos}

\affiliation[a]{Institut für Theoretische Physik, Goethe Universität, Max-von-Laue-Str. 1, 60438 Frankfurt, Germany}
\affiliation[b]{Institute of Physics, National Yang Ming Chiao Tung University, 30010 Hsinchu, Taiwan}
\affiliation[c]{School of Mathematics, Trinity College Dublin, Dublin 2, Ireland}
\affiliation[d]{Jülich Supercomputing Centre, Forschungszentrum Jülich GmbH, 52428 Jülich, Germany}
\affiliation[e]{$CP^3$-Origins, University of Southern Denmark, 5230 Odense, Denmark}
\affiliation[f]{FRIB \& Physics Department, Michigan State University, East Lansing, MI 48824, USA}
\affiliation[g]{Nuclear Science Division, Lawrence Berkeley National Laboratory, Berkeley, CA 94720, USA}
\affiliation[h]{Aix Marseille Univ, Université de Toulon, CNRS, CPT, Marseille, France.}

\emailAdd{g.pederiva@fz-juelich.de}

\abstract{
  The OpenLat initiative presents its results of lattice QCD simulations using Stabilized Wilson Fermions (SWF) using 2+1 quark flavors. Focusing on the $\mathrm{SU}(3)$ flavor symmetric point $m_\pi=m_K=412$ MeV, four different lattice spacings ($a=0.064,0.077,0.094,0.12$ fm) are used to perform the continuum limit to study cutoff effects.

  We present results on light hadron masses; for the determination we use a Bayesian analysis framework with constraints and model averaging to minimize the bias in the analysis.
}

\FullConference{
  The 39th International Symposium on Lattice Field Theory (Lattice2022),\\
  8-13 August, 2022 \\
  Bonn, Germany
}

\begin{document}
\maketitle

\section{Introduction}
In recent years there has been a growing interest in Stabilized Wilson Fermions (SWF), with the OpenLat initiative playing a crucial role in their development. The SWF is a set of both algorithmic and analytical improvements over standard Wilson-Clover fermions designed to overcome some of their issues, for example SWF allow for simulations at coarse lattice spacing even at small pion masses~\cite{francis_master-field_2020}. The SWF package includes the use of the Stochastic Molecular Dynamics (SMD) algorithm instead of the HMC and the exponential clover action among other things.

The study of SWF has been initiated in the context of master-field simulations \cite{francis_master-field_2020} and since then more simulations where new encouraging results have been produced~\cite{fritzsch_master-field_2022,ce_approaching_2022}. The open lattice initiative (OpenLat), on the other hand, was founded~\cite{cuteri_open_2021} with the goal of generating state-of-the-art QCD ensembles using SWF and share them according to the principles of the open science philosophy with all the LQCD community.

One of the first results that is planned for the newly generated ensembles is that of the light hadron spectrum as benchmark and in order to assess the discretization effects. The spectrum is
determined from the two-point correlation functions
using a Bayesian analysis framework and model averaging.
A principle goal with this strategy is to reduce potential human bias in the analysis
and to establish solid benchmarks.

\section{OpenLat Ensembles}
\begin{figure}[htbp!]
    \centering
    \includegraphics[width=0.49\textwidth]{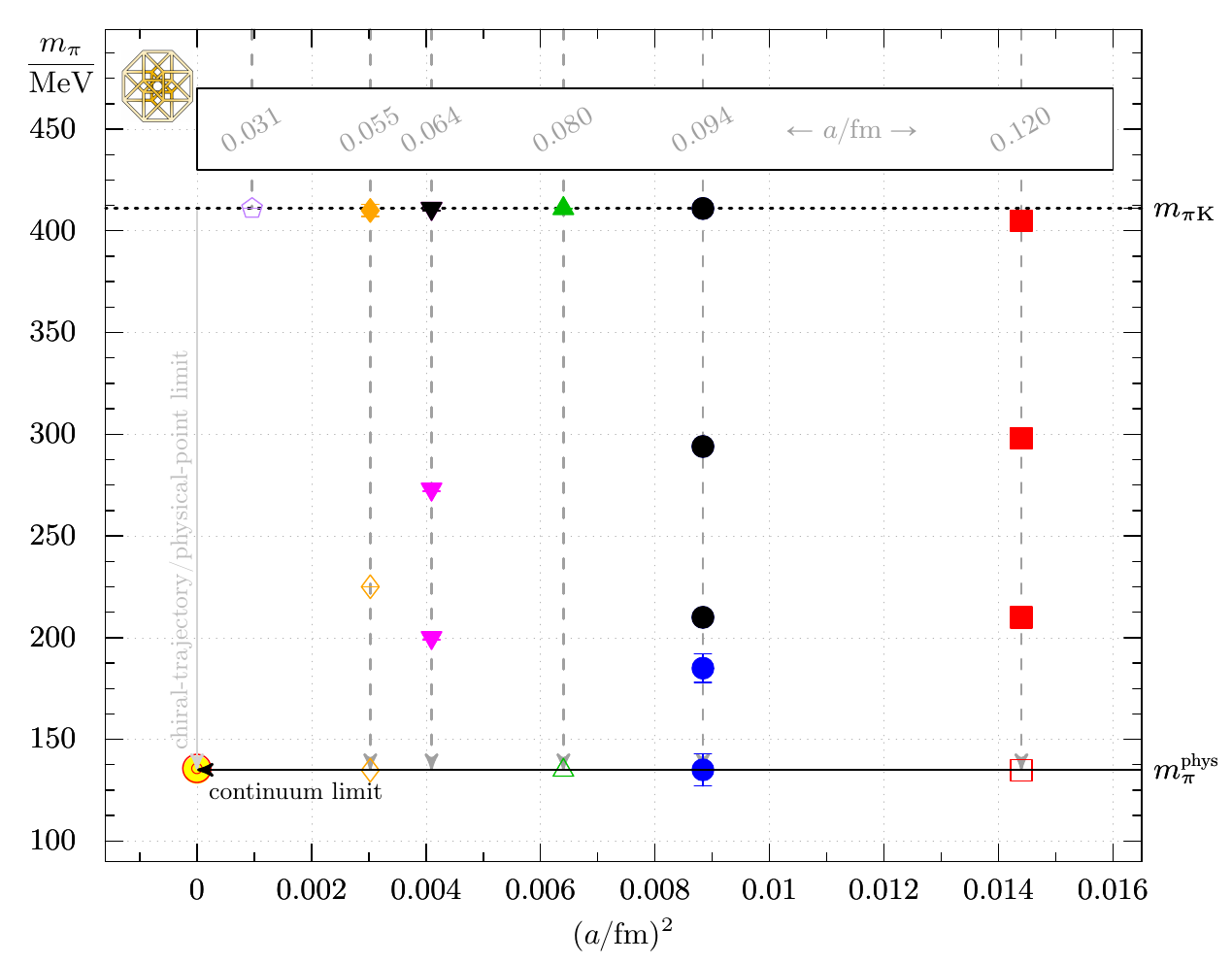}
    \includegraphics[width=0.49\textwidth]{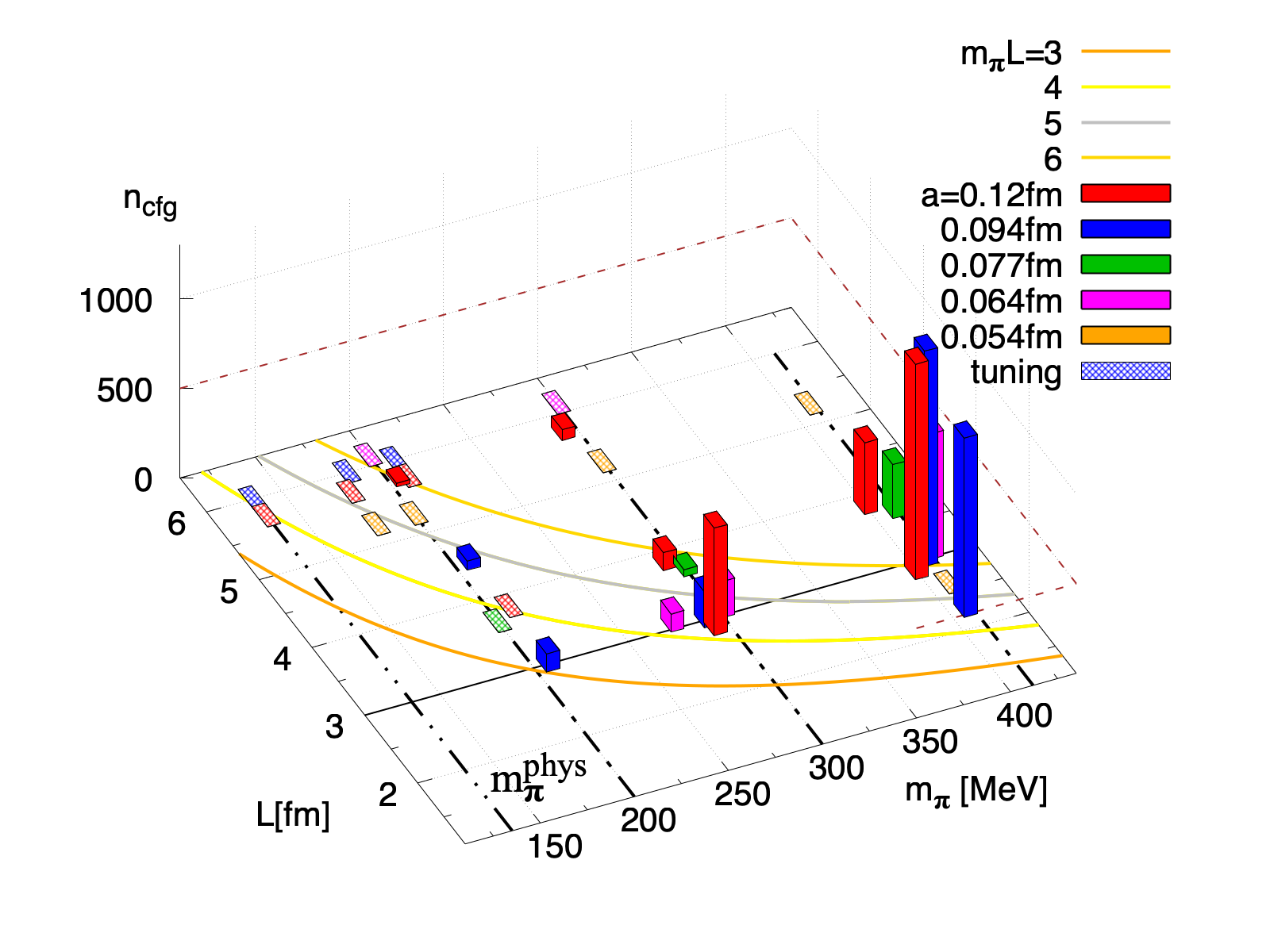}
    \caption{Overview of the OpenLat ensembles presented here as parametrized by the lattice spacing $a$ and $m_\pi L$. The height of the bars in the histograms represents the number of independent configurations that have been generated. Shaded boxes are ensembles in the tuning state~\cite{francis_properties_2022,francis_gauge_2022}}~\label{fig:ens-status}
\end{figure}

The ensemble generation follows a three stages approach, at the end of which a reference publication will be released~\cite{francis_gauge_2022}, along with the configurations and their respective metadata used in the publication.
The metadata consists of all the observables used for validation during the generation. We also want to include a study of the light hadron spectrum to be used as reference and benchmark for the discretization effects of the SWF ensembles. In Figure~\ref{fig:ens-status} the parameter space for the ensemble that are being produced and planned is reported, together with their current status.

For this study of hadron spectroscopy, the ensembles collected in \Cref{tab:swf-ens} were used\footnote{More details about the generation process can be found in the proceeding of this conference for the talk by A. Francis~\cite{francis_gauge_2022}}.

\begin{table}[htb!]
    \centering
    \begin{tabular}{ccccccc}
        \hline 
        Ensemble           & $L^3\times T$   & $t_0/a^2$      & $a$ [fm] & $N_{cfg}$ & $N_{srcs}$ \\\hline 
        \texttt{a12m412mL5.9}   & $24^3\times 96$ & $1.48548(64) $ & $0.12$   & $600$     & $100$      \\
        \texttt{a094m412mL4.6} & $24^3\times 96$ & $2.4482(50) $  & $0.094$  & $500$     & $100$      \\
        \texttt{a094m412mL6.2}  & $32^3\times 96$ & $2.43979(89) $ & $0.094$  & $1200$    & $100$      \\
        \texttt{a077m412mL7.6}  & $48^3\times 96$ & $3.6198(30)  $ & $0.077$  & $900$     & $100$      \\
        \texttt{a064m412mL6.3}  & $48^3\times 96$ & $5.2588(46)  $ & $0.064$  & $1000$    & $100$      \\
    \end{tabular}
    \caption{SWF Ensembles used for the analysis. We note that the ensemble
    key denotes the approximate lattice spacing, pion mass and value of $m_\pi L$ respectively.
    $N_{cfg}$ represents the number of gauge field configurations analyzed, but not all of them are to be considered independent. $N_{srcs}$ is the number of stochastic sources per configuration used to compute the averaged hadron correlators.}~\label{tab:swf-ens}
\end{table}

\section{A Bayesian Analysis Framework for Hadron Masses}~\label{ssec:bayes}
One issue with multi-state fitting for hadron correlators is the numerical stability of the algorithm coming from the non-linearity of the fit function.
This poses several challenges in fitting the correlation functions with a sum of exponentials, including
(i) A naive approach necessarily introduces an arbitrary cut in the number of states, which can bias the determination of the ground state parameters;
(ii) The minimizer can be very sensitive to the initial guess values of the parameters and get stuck in local minimum of the $\chi^2$.
In our determination of the hadron spectrum we use an unbiased method based on constrained Bayesian fitting and model averaging to minimize the bias in the results.

\subsection{Bayesian Fitting with Constraints}
A solution for improving the stability of multi-state fits can be the introduction of
Bayesian constraints with reasonable estimates for the priors
, as first introduced by Lepage et al.~in~\cite{lepage_constrained_2002} for the case of correlators from LQCD.
In our analysis we computed the correlators
with a covariant Gaussian smeared source and the same smearing at the sink as well as a point sink, which we denote as $SS$ and $PS$.
The functional form of the multi-exponential to be fitted is then:
\begin{align}
C_{SS}(t; Z_{S,n}, E_n) &=
    \sum_{n=0}^{n=N_s} Z_{S,n}^2 e^{-E_n t}\, ,
\nonumber\\
C_{PS}(t; Z_{P,n}, Z_{S,n}, E_n) &=
    \sum_{n=0}^{n=N_s} Z_{S,n}Z_{P,n} e^{-E_n t}\, ,
\end{align}
where $N_s$ is the number of states to be fit, $Z_{P/S}$ are the amplitudes for the point and smeared source and sinks and $E_n$ are the energy levels. The fit is constrained by the introduction of priors for the fit parameters, denoted with a tilde, and their uncertainties chosen as follows:
\begin{align}
    \tilde{Z}_{P,0},         &\quad \text{determined from $z^P_{\mathrm{eff}}$ data}~~~ & \tilde{Z}_{P,i} = \tilde{Z}_{P,0} ~~                                                                         & \text{for}~i=1\dots N_s  \\\nonumber
    \tilde{Z}_{S,0}          &\quad \text{determined from $z^S_{\mathrm{eff}}$ data}~~~ & \tilde{Z}_{S,1} = \tilde{Z}_{S,0};  \tilde{Z}_{S,i}=\frac{\tilde{Z}_{S,0}}{2}~~                              & \text{for}~i=2\dots N_s  \\\nonumber
    \sigma_{\tilde{Z}_{P,0}} &\quad \text{determined from $z^P_{\mathrm{eff}}$ data}~~~ & \sigma_{\tilde{Z}_{P,i}} = 2\sigma_{\tilde{Z}_{P,0}} ~~                                                      & \text{for}~i=1\dots N_s  \\\nonumber
    \sigma_{\tilde{Z}_{S,0}} &\quad \text{determined from $z^S_{\mathrm{eff}}$ data}~~~ & \sigma_{\tilde{Z}_{S,1}} = 2\sigma_{\tilde{Z}_{S,0}};  \sigma_{\tilde{Z}_{S,i}} = \sigma_{\tilde{Z}_{S,0}}~~ & \text{for}~i=2\dots N_s.
\end{align}
The effective overlap data is given by
\begin{align}
&z_S^{\mathrm{eff}}(t) = \left[ e^{m_{\mathrm{eff}}(t) t} C_{SS}(t) \right]^{1/2}\, ,&
&z_P^{\mathrm{eff}}(t) = \frac{e^{m_{\mathrm{eff}}(t) t} C_{PS}(t)}{z_S^{\mathrm{eff}}(t)}\, ,&
\end{align}
and the effective mass is the average of that determined from the SS and PS correlators
\begin{equation}
m_{\mathrm{eff}}(t) = \ln \left( \frac{C_{XS}(t)}{C_{XS}(t+1)} \right)\, ,
\quad X=\{S,P\}\, .
\end{equation}

\begin{figure}[htbp!]
    \centering
    \includegraphics[width=0.32\textwidth]{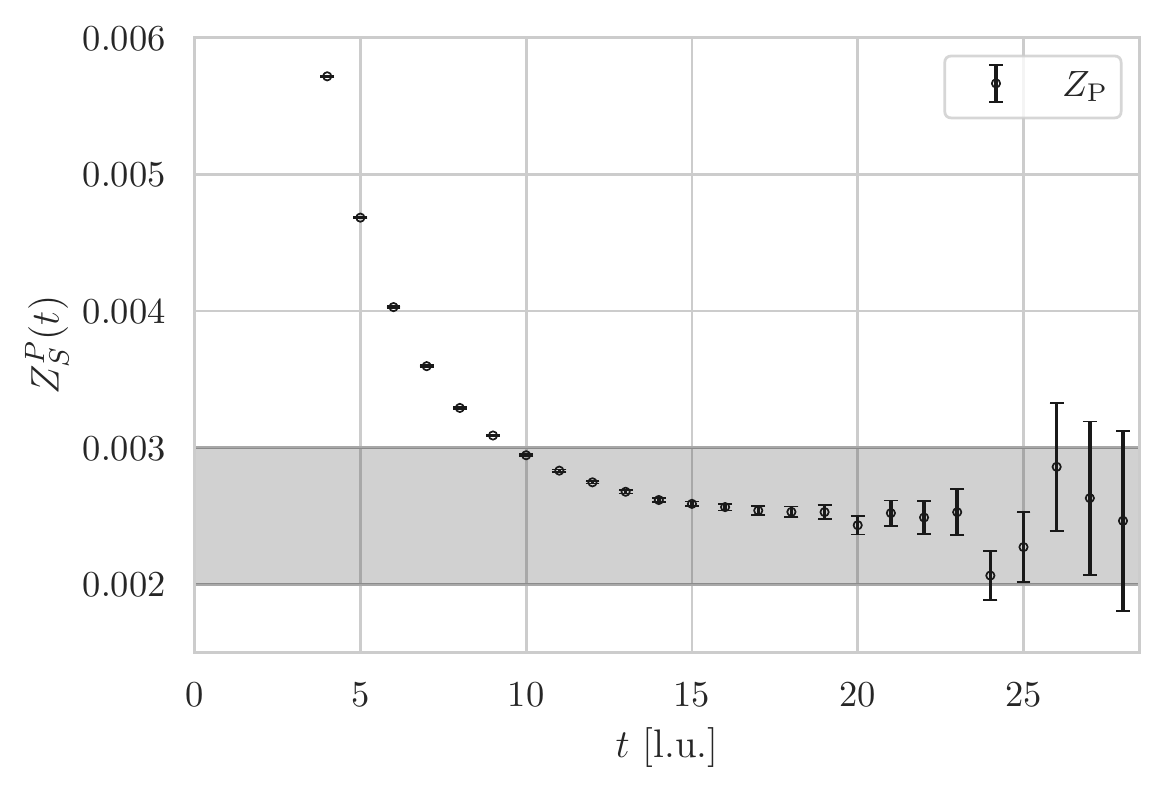}
    \includegraphics[width=0.32\textwidth]{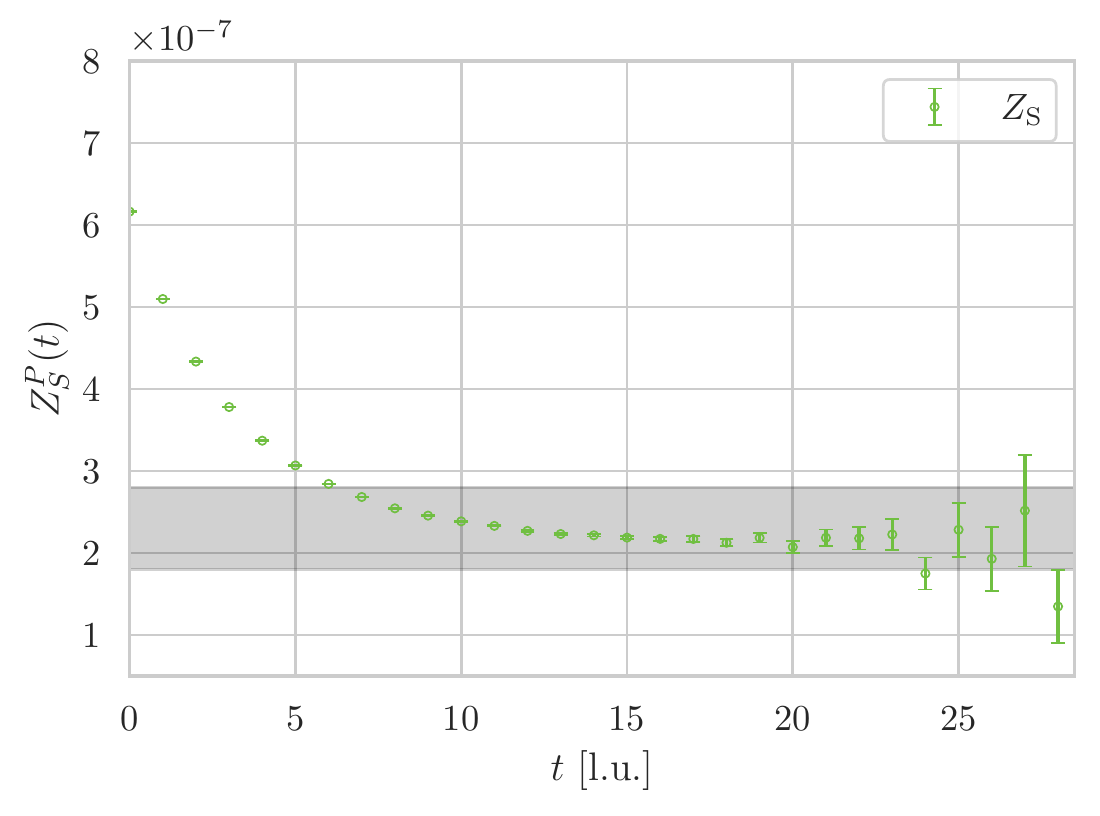}
    \includegraphics[width=0.32\textwidth]{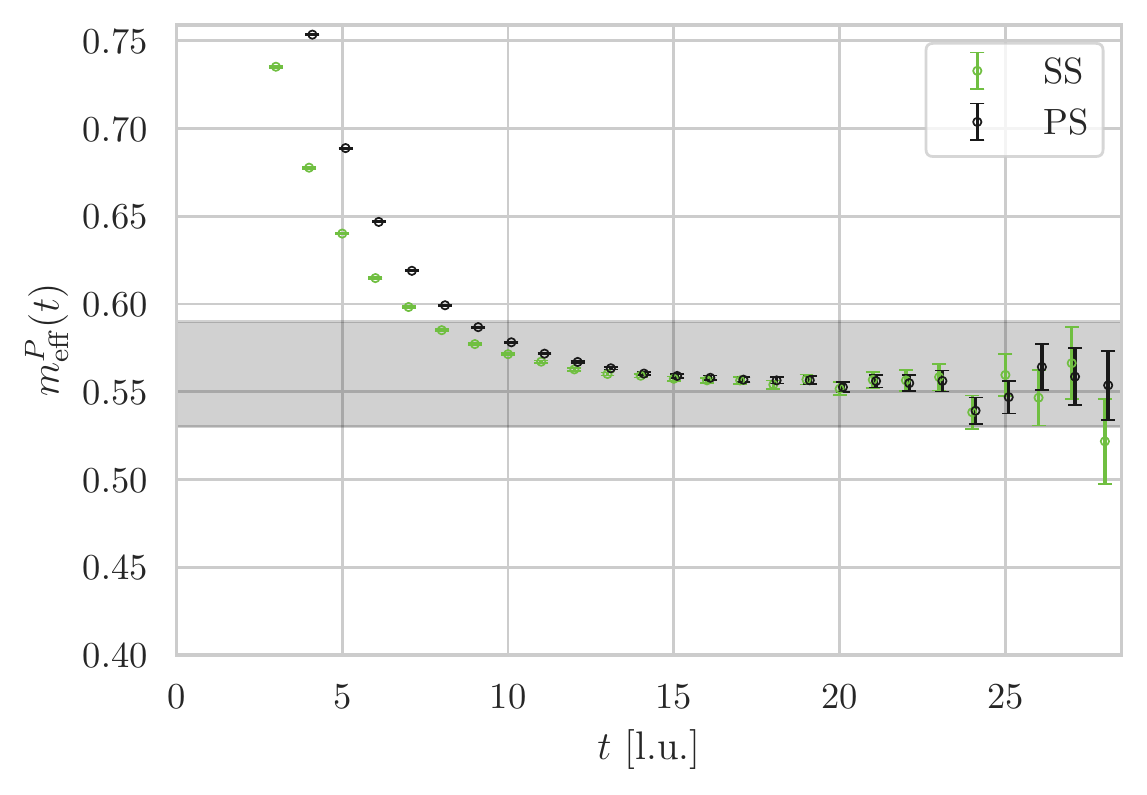}
    \caption{Amplitudes for the point-smeared (left) and smeared-smeared (right) correlators and the respective prior choices. The data is from the $a094m412$ ensemble, with 100 stochastic sources per configuration.}~\label{fig:priors}
\end{figure}

The use of the data to estimate the priors is potentially problematic.  In order to prevent biassing the determination of the posteriors, we use the effective mass and effective overlap factor processed data and chose uncertainties for the priors that are approximately 10 times larger than the anticipated posterior uncertainty~\cite{Miller:2020evg}, see figure~\ref{fig:priors}.
In order to estimate the priors for the excited state overlap factors, we use
the expectation that smearing should reduce the overlap with higher excited states, thus $\tilde{Z}_{S,n}$ is considered to be suppressed for high $n$.
For the excited state $Z_{P,n}$ priors, we set the prior width to be twice as large as for the ground state with the mean value estimated to be the same as for the ground state.
For the excited state energies, we use a model where the gap from the $i$ to the $i+1$ state is priored to be $2m_\pi$ with a log-normal distribution 
that reduces to $m_\pi$ at one-sigma.
The use of log-normal priors forces the excited state energy gaps to be positive-definite and thus the posterior energies remain ordered~\cite{lepage_constrained_2002}.
The choice of $\tilde{E}_{i+1}-\tilde{E}_i = 2m_\pi$ is phenomenologically motivated.  For values of $m_\pi L\approx4$, the single soft-pion excitation in a p-wave has a similar energy.
\begin{align}
    \tilde{E}_0,            &\quad \text{determined from $m_{\mathrm{eff}}(t)$ data}~~~ & \ln(\tilde{E}_{i} - \tilde{E}_{i-1}) = \ln(2m_\pi) ~~ & \text{for}~i=1\dots N_s  \\\nonumber
    \sigma_{\tilde{E}_{0}}, &\quad \text{determined from $m_{\mathrm{eff}}(t)$ data}~~~ & \sigma_{\tilde{E}_{i}} = \sigma_{\tilde{E}_{0}}       & \text{for}~i=1\dots N_s.
\end{align}
The values for $\sigma_{\tilde{E}_{0}}, \sigma_{\tilde{Z}_{P,0}}$ and $\sigma_{\tilde{Z}_{S,0}}$ are chosen to be $~10$ times the expected value from the data (the shaded regions in Figure~\ref{fig:priors}).

The $\chi^2$ function to minimize is modified to include the contributions from the priors by defining the augmented $\chi^2$:
\begin{align}\label{eq:chi-augmented}
    \chi^2 \rightarrow \chi^2_{\mathrm{aug}} &= \chi^2 + \chi^2_{\mathrm{prior}}
\nonumber\\
    \text{with}\quad & \chi^2_{\mathrm{prior}} =
    \sum_n^{N_s} \left[
    \frac{(Z_{P,n} - \tilde{Z}_{P,n})^2}{\sigma_{\tilde{Z}_{P,n}}^2}
    + \frac{(Z_{S,n} - \tilde{Z}_{S,n})^2}{\sigma_{\tilde{Z}_{S,n}}^2}
    + \frac{(E_n - \tilde{E}_n)^2}{\sigma_{\tilde{E}_n}^2}
    \right]\, .
\end{align}

\subsection{Bayesian Model Averaging}
To further determine an unbiased value for the mass of the hadrons, we use a Bayesian model averaging procedure based on the Akaike Information Criterion (AIC) that has been recently proposed in~\cite{jay_bayesian_2021} for the case of lattice correlator data with data windowing selection. This procedure further eliminates the need to ``manually'' check for convergence in the constrained fitting procedure as the number of states increases. In turn, it defines a weight for every fit result, parametrized by the pair $t_{\min},N_s$ that is used to determine the weighted average of the fits.

Given a fit result with parameters $M = {A_1,\dots A_n, E_1,\dots E_n}$ the AIC is:
\begin{equation}
    \mathrm{AIC}_M = -2\log(\mathrm{pr}(M)) + \chi^2_{\mathrm{aug}} + 2k + 2n
    \label{eq:aic}
\end{equation}
where $\mathrm{pr}(M)$ is the model likelihood function of the model, $k$ is the number of fit parameters, including priors, and $n$ is the number of excluded points from the fit, i.e.\ it depends on changes of $t_{\min}$. The minimum value of the AIC among all models is used to determine the relative likelihood of a given model $M_i$ and some data $D$ as:
\begin{equation}
    \mathrm{pr}(M_i|D) = \frac{\exp(\mathrm{AIC}_{\min} - \mathrm{AIC}_i)}{\sum_j \exp(\mathrm{AIC}_{\min} - \mathrm{AIC}_j)}
\end{equation}
where the denominator is just to fix the normalization to allow a probabilistic interpretation of the quantity. The relative likelihood is then used as weight for computing the weighted sum of a fit parameter $\rho \in M$. The unbiased estimator for the parameter is then its weighted average and its uncertainty is given by:
\begin{equation}
    \sigma_\rho = \sum_i \sigma_{\rho,i} \mathrm{pr}(M_i|D) + \sum_i \langle \rho \rangle_i^2 \mathrm{pr}(M_i|D) - {\left(  \sum_i \langle \rho \rangle_i \mathrm{pr}(M_i|D) \right)} ^2
\end{equation}

There are a few considerations to make regarding this procedure. The first and most important is that it requires very little input, in particular it requires making an educated guess for the priors and attaching to their values a reasonably large uncertainty such that they do not become dominant in the fit; this greatly reduces the bias of the fit. Another interesting feature that can be immediately inferred from \Cref{eq:aic} is that this method favors models that are simpler, i.e.\ fit functions with fewer number of states. This sort of built-in Occam's razor is very useful to gain intuition on how many states can be determined from the given set of data a posteriori. At the same time, \Cref{eq:aic} tells us that this procedure favors models with larger fitting ranges, hence more data, which is a feature that we implicitly desire.

\section{Numerical Results}~\label{sec:spec-results}
The calculations of the hadron correlators were performed using the \texttt{lalibe}~\cite{gambhir_code_nodate} software, which builds on top of \texttt{Chroma}~\cite{edwards_chroma_2005}. The chosen smearing parameters are $N_{\mathrm{smear}} = 32$ and $\sigma = 3.86$, where $\sigma$, following the notation used in~\cite{gusken_study_1990}, is $\sqrt{4 \alpha N_{\mathrm{smear}}}$ and represents a dimensionless smearing radius.

\begin{figure}[htbp!]
    \centering
    \includegraphics[width=0.8\textwidth]{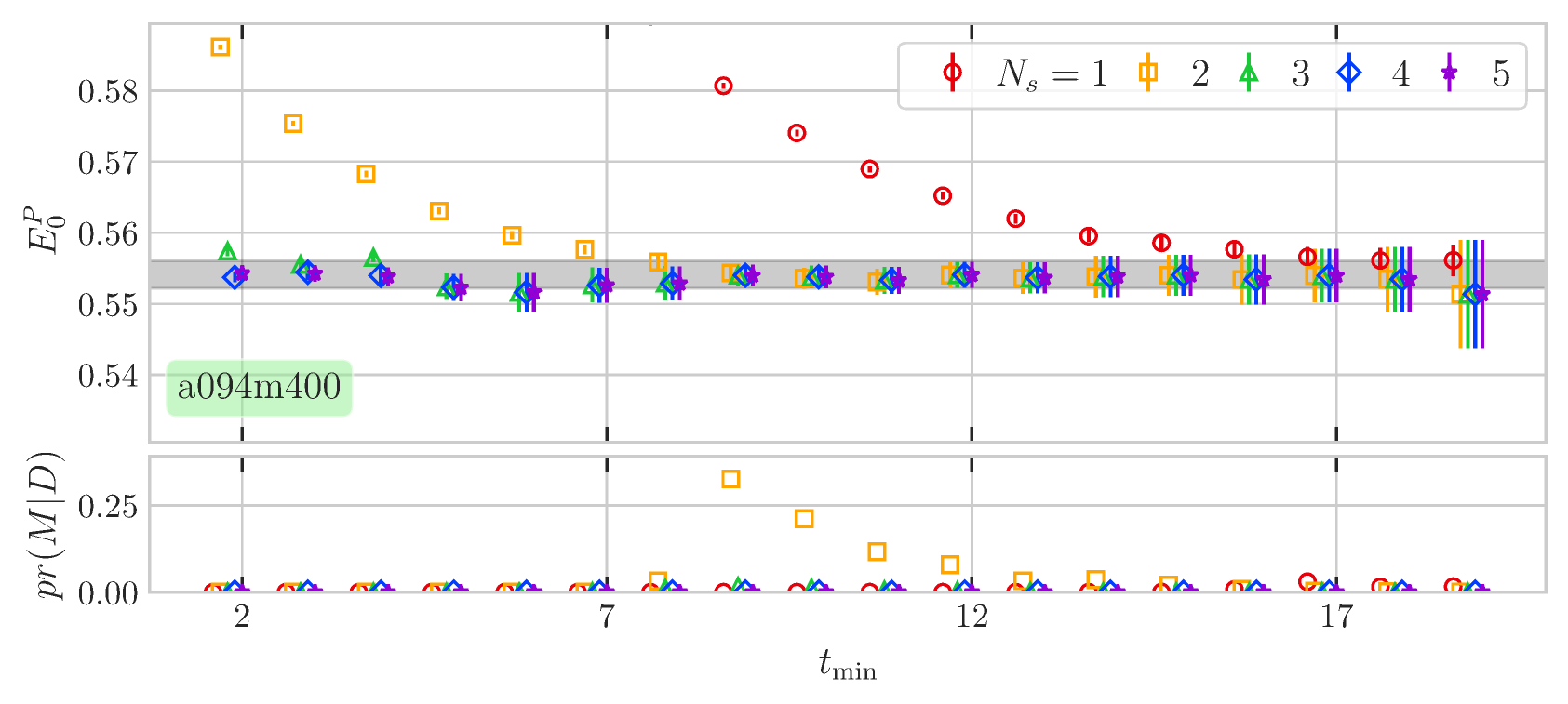}
    \caption{Fit results for the proton mass for the \texttt{a094m412} ensemble. }~\label{fig:fit-proton}
\end{figure}

As outlined in the previous section, a scan over a range of $t_{\min}$ is made, while $t_{\max}$ is fixed to a value where it does not affect the fit results due to the low signal-to-noise. All the results for $N_S=1,\dots,5$ are then combined using model averaging. In \Cref{fig:fit-proton} we show our results for the \texttt{a094m412mL6.2} ensemble for the proton as an example.

As discussed in~\cite{lepage_constrained_2002}, all fits should converge for large enough $N_S$ given that there is enough information in the $t$ range to fix the ground state. This is in fact the case: for low $N_S$ we observe that the ground state estimate converges to the results coming from fits with more states for large enough $t_{\min}$, where there is no information on the higher states. The height of the top panel is the in \Cref{fig:fit-proton} is set to be the width of the prior of the ground state energy. The model averaging result and error, shown as the gray band in the plot, is indeed compatible with all the converged fits and has, as expected, an uncertainty of the same order of the individual fits.

The lower panel of \Cref{fig:fit-proton} shows the weights of the different fits as they enter the average. One observes that our procedure based on the AIC selects the fits with low $N_S$ and low $t_{\min}$. In the case of the proton shown above most of the contribution comes from the $N_S=2$ fits.

\begin{figure}[htbp!]
    \centering
    \includegraphics[width=0.6\textwidth]{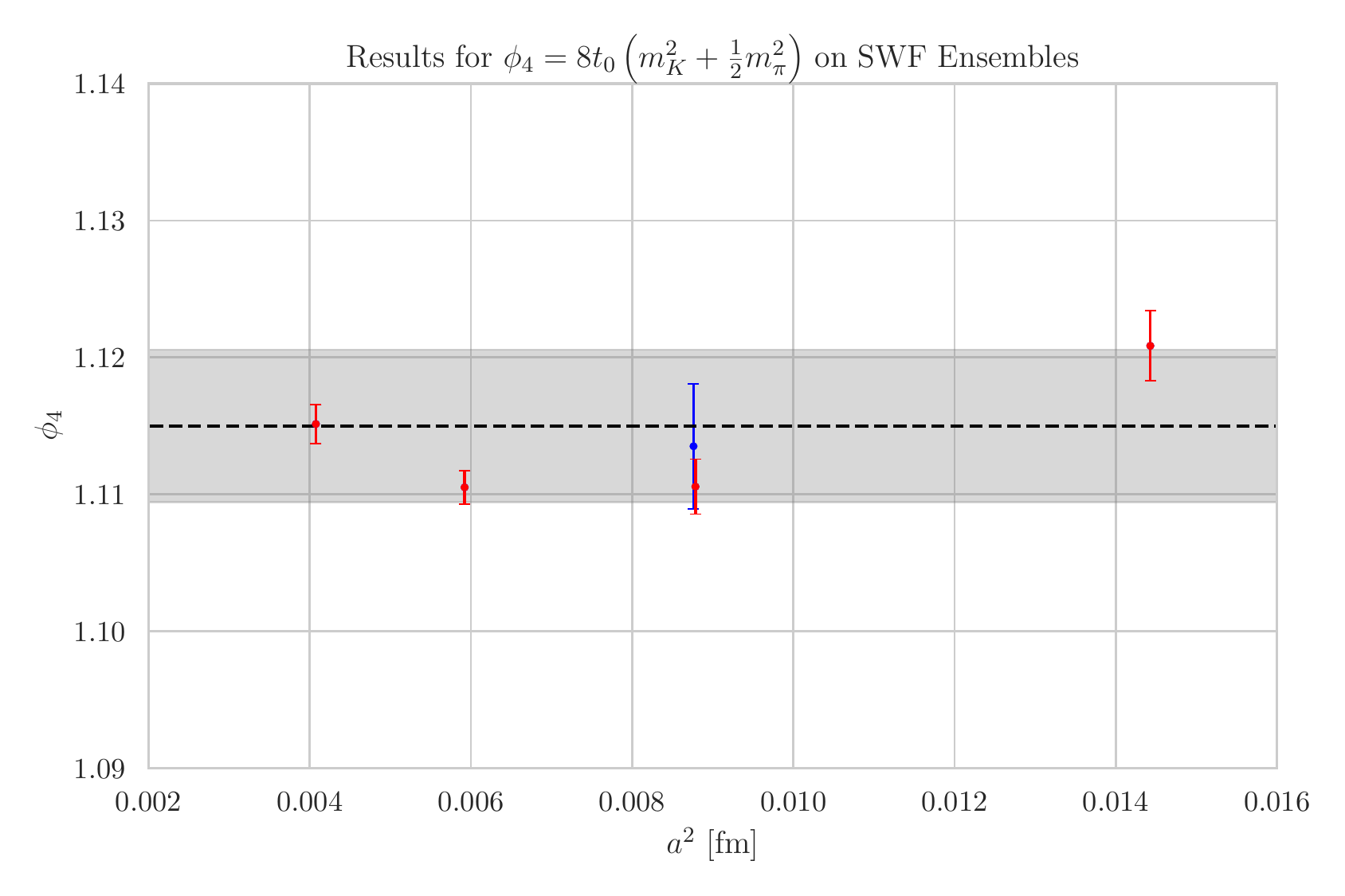}
    \caption{Values of $\phi_4$ for SWF at the $\mathrm{SU}(3)$ flavor-symmetric point.}~\label{fig:phi4}
\end{figure}

The fitting procedure and model averaging is used on all ensembles for the pion, proton and $\Omega$ baryon, which are the only hadrons at the $\mathrm{SU}(3)$ flavor symmetric point. First, in \Cref{fig:phi4} we show $\phi_4 = 8t_0(m_K^2 + \frac{1}{2}m_\pi^2)$, to check the mass tunings of the ensembles, according to the imposed quality criteria the value of $\phi_4$ should be within the gray band at the $1\sigma$ level, which is indeed the case\footnote{The authors would like to point out that this figure is different from the one shown at during the conference presentation due to the finding of a mistake in one of the parameters.}.

\begin{figure}[htbp!]
    \centering
    \includegraphics[width=0.49\textwidth]{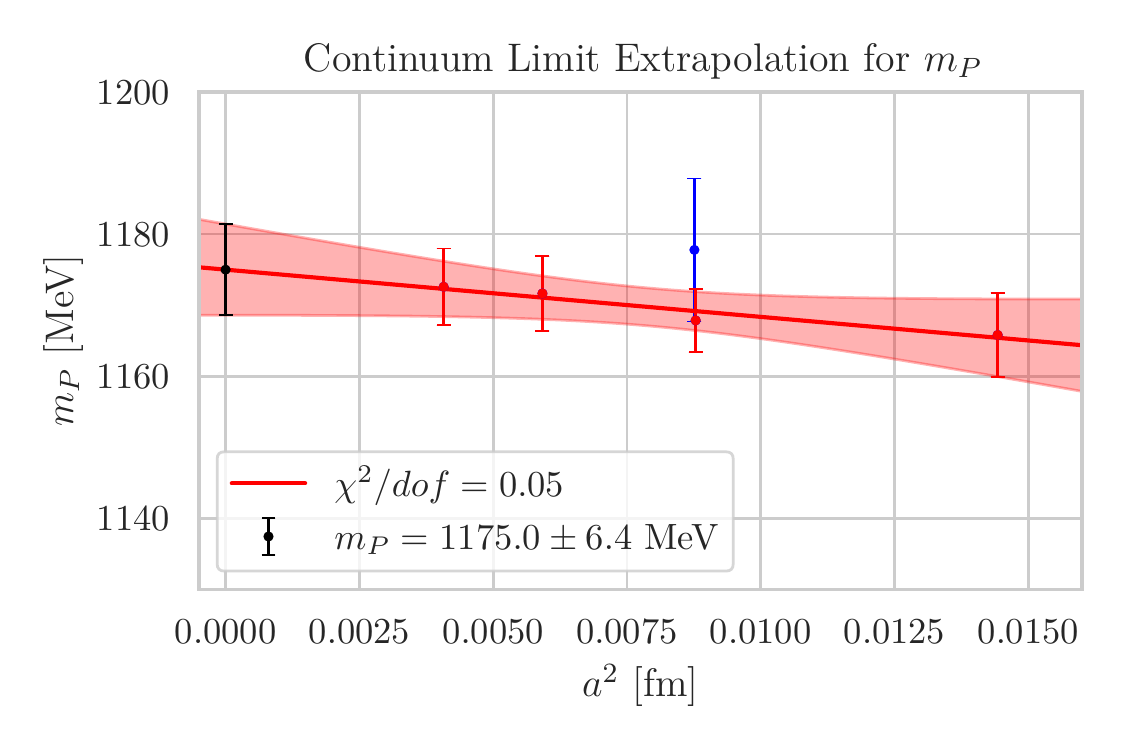}
    \includegraphics[width=0.49\textwidth]{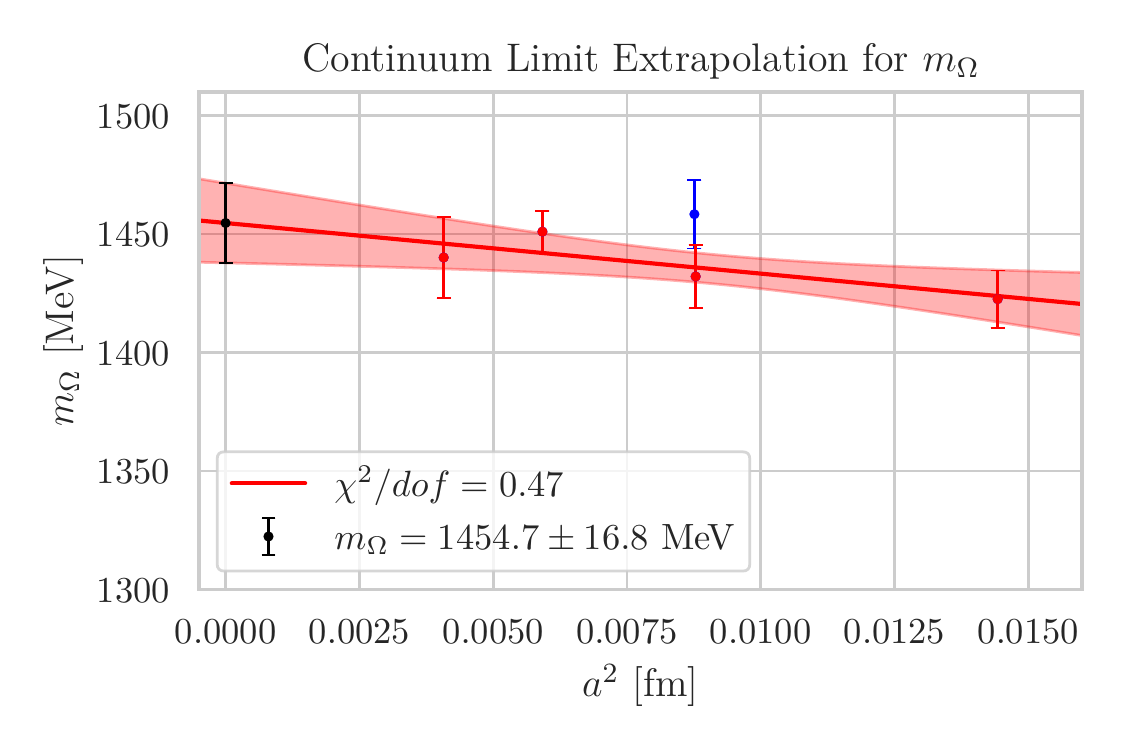}
    \caption{Continuum limit extrapolation for the proton (left) and $\Omega$ (right) masses for SWF at the $\mathrm{SU}(3)$ flavor-symmetric point.}~\label{fig:cont-proton}
\end{figure}

In \Cref{fig:cont-proton}, we show aggregated results for the proton and $\Omega$ mass as a function of $a^2$. One notices that the given SWF ensembles show very small cutoff effects for hadronic observables. The continuum extrapolations for the masses of the proton and $\Omega$ are $m_p = 1175.0 \pm 6.4$ MeV and $m_\Omega = 1454.7 \pm 16.8$.

\section{Outlook for Studies with SWF}
The results for the scaling properties of the hadron spectrum are very encouraging and show the validity and quality of the new SWF ensembles that have been generated by the OpenLat initiative. The analysis presented here is an ongoing work that will be repeated for every ensemble produced by the collaboration. The results will be included in the publication accompanying the public release of the ensembles in the future.

\section*{Acknowledgements}
The authors acknowledge support from the HPC computing centers hpc-qcd (CERN), HPE Apollo Hawk (HLRS) under grant stabwf/44185, 
Piz Daint (CSCS), Occigen (CINES), Jean-Zay (IDRIS) and Ir\`ene-Joliot-Curie (TGCC) under projects (2020,2021,2022)-A0080511504, (2020, 2021, 2022)-A0080502271 by GENCI and PRACE project 2021250098. This work also used the DiRAC Extreme Scaling service at the University of Edinburgh, operated by the Edinburgh Parallel Computing Centre on behalf of the STFC DiRAC HPC Facility. DiRAC is part of the UK National e-Infrastructure.
This work also used resources of the National Energy Research Scientific Computing Center (NERSC), a U.S. Department of Energy Office of Science User Facility located at Lawrence Berkeley National Laboratory, operated under Contract No. DE-AC02-05CH11231 using NERSC awards NP-ERCAP0020427, NP-ERCAP0017010 and NP-ERCAP0014740.
This work also used the Frontera HPC system operated by the Texas Advanced Computing Center~\cite{frontera}.
AS acknowledges funding support under the National Science Foundation grant PHY-2209185. AF acknowledges support under the Ministry of Science and Technology Taiwan grant 111-2112-M-A49-018-MY2.

\bibliographystyle{JHEP}
\bibliography{pederiva-lattice-2022}

\end{document}